\documentclass{article}

\input epsf

\def\DR{\rm I\kern-1.45pt\rm R}
\def\DC{\kern2pt {\hbox{\sqi I}}\kern-4.2pt\rm C}

\newcommand{\ba}{\begin{array}}
\newcommand{\ea}{\end{array}}
\newcommand{\be}{\begin{equation}}
\newcommand{\ee}{\end{equation}}
\newcommand{\bea}{\begin{eqnarray}}
\newcommand{\eea}{\end{eqnarray}}
\newcommand{\bi}{\begin{itemize}}
\newcommand{\ei}{\end{itemize}}

\usepackage{amscd,amsmath,amssymb}
\begin{document}\begin{center}
{\bf \Large  Multi-center MICZ-Kepler systems}\\
\vspace{0.5 cm} {\large  Armen Nersessian$^{1,2}$ and Vadim
Ohanyan$^{1,3}$ }
\end{center}
\noindent
 $\;^1$ {\it Yerevan State University , 1 Alex Manoogian
St., Yerevan, 375025
 Armenia}\\
{\it
 \&
  Yerevan Physics Institute, 2 Alikhanian Brothers St.,
 Yerevan,
  375036,
    Armenia}\\
    $\;^2$ {\it Artsakh State University, 3 Mkhitar Gosh St.,
  Stepanakert,
Armenia}\\
${}^{3}$ {\it Russian-Armenian University, 123 Hovsep Emin St,
 Yerevan, 375051,
 Armenia}\\
 {\sl E-mails:
  arnerses@yerphi.am, ohanyan@yerphi.am}
\begin{abstract}
We present the classical solutions of the   two-center MICZ-Kepler
and MICZ-Kepler-Stark systems.
 Then we suggest the  model of  multi-center
 MICZ-Kepler system on the curved spaces equipped with $so(3)$-invariant conformal flat metrics.
\end{abstract}

\section{Introduction}
As is known, the Kepler system plays a special role among the
finite-dimensional classical and quantum integrable systems. On
the one hand, it has a number of physical applications, on the
other hand, this system is one of the key components of hidden
symmetry concept\cite{lan}.  Various generalizations of this
system were studied during last century. Among them are the Kepler
problems on spheres and hyperboloids \cite{sch41} as well as their
higher dimensional counterparts. Further important generalization
of the Kepler problem was constructed in the works of Zwanziger,
and McIntosh and Cisneros \cite{zwa68}. They replaced the Coulomb
center of the system by the Dirac dyon (electrically charged Dirac
monopole \cite{dir31}), and added to the initial Coulomb potential
the specific centrifugal term $s^2/2 \mu r^2$, where $s=eg/c$ is
the so-called monopole number ($e$ is the electrical charge of the
probe particle and $g$ is the magnetic charge of monopole), $\mu$
is a particle mass. As a result they obtained the integrable
system which is very similar to the underlaying Kepler one. It is
presently known as a MICZ-Kepler system. For instance, MICZ-Kepler
system inherent, besides the rotational symmetry, the hidden
symmetry of the Kepler system given by the Runge-Lenz vector.
Similar to the case of pure Coulomb system, the classical
trajectories of the MICZ-Kepler system are second order curves.
The quantum mechanical properties of Coulomb and MICZ-systems have
also much in common. For example, the only difference in their
energy spectra is the lift of the low bound of the orbital quantum
number from $0$ to $|s|$.

The MICZ-Kepler system describes the motion of the electrically
charged scalar  particle  in the field of the Dirac dyons. The
origin of the additional centrifugal term $\frac{s^2}{2\mu r^2}$
can be  also understood in this context. In the presence of
monopole magnetic field  the angular momentum of the system get
additional, spin-like term, ${\bf J}={\bf r}\times {\bf\pi} -
s{\bf r}/r$, where ${\bf \pi}$ is dynamical momentum containing
the vector potential of the monopole magnetic field. The magnetic
field of Dirac dyon  is ${\bf B}=g{\bf r}/{r^3}$. On the other
hand, there is linear relation between orbital and magnetic
momenta of single particle,
   ${\mathbf{{\mathcal{M}}}}=\frac{e}{2 \mu c
  }{\mathbf{J}}$.
Hence, the  interaction energy of this magnetic momentum with
magnetic field ${\bf B}$ is given by the expression:
  \be
  U_B=-{\mathbf{{\mathcal{M}}}}{\mathbf{B}}=-\frac{eg}{2 \mu
  c}{\mathbf{J}}\frac{{\bf r}}{r^3}=\frac{s^2}{2 \mu r^2}, \label{UB}
  \ee
 i.e. coincides with the centrifugal term in the MICZ-Kepler system \cite{terant}.
 Though, the additional part of the orbital
  momentum proportional to $s$ should rather be assigned to the
  electro-magnetic field than to the particle, this non-correct interpretation
  nevertheless leads to the correct expression.
Let us also mention, that MICZ-Kepler system also describes the
relative motion of two Dirac dyons with electric and magnetic
charges $(e_1, g_1)$ and $(e_2, g_2)$ with ${e_1g_2-e_2g_1}={ c}s
$. Similar to above mentioned case, the centrifugal term $s^2/2\mu
r^2$ could be interpreted as the interaction energy of the induced
dipole momentum with the dyon electric field plus interaction
energy of the induced magnetic moment with magnetic field (here
$\mu$ is the reduced mass).

 The analogy between MICZ-Kepler and
Coulomb systems has elegant explanation in terms of
four-dimensional space: both systems can be obtained from the
Hamiltonian of four-dimensional oscillator via the reduction by
$U(1)$ group action \cite{micred}. In the same way, one can
construct the five-dimensional analog of the MICZ-Kepler problem,
reducing the eight-dimensional oscillator by $SU(2)$ group action
\cite{su2}. In this case, the $SU(2)$ Yang monopole appears in the
system instead of Dirac monopole.
 In two-dimensional case  one can  also construct MICZ-Kepler like system from the oscillator:
 for this purpose the two-dimensional (quantum) oscillator should be reduce  by the
action of the  parity ($Z_2$) group.
  The emerged two-dimensional MICZ-Kepler system is distinguished by the presence
  of  magnetic flux  which provide the system with spin $1/2$ ($\mathbb{Z}_2$ monopole) \cite{ntt}.
The reduction procedure mentioned above have very deep and close
relation with the  Hopf maps (see, e.g.
\cite{lnp}).
 For the detailed review of the quantum mechanical aspects of these issues the reader may see the Ref. \cite{book}.
The  MICZ-Kepler systems on the three-dimensional spheres
\cite{kurochkin}, on  two-, three- ,  and five-dimensional
two-sheet hyperboloids \cite{pogos} as well as their
generalization to arbitrary dimensions \cite{meng} are also
distinguished by their close similarity to the corresponding
Coulomb systems. For example, the only difference between the
quantum-mechanical properties of Coulomb and the corresponding
MICZ-Kepler systems consists in the lift of the range of the total
angular momentum, which in its turn leads to the degeneracy of the
ground state. Indeed, any three-dimensional rotationally
invariant system (without monopoles) will preserve its main
properties if incorporation of the the Dirac monopole in the
center will be supplied by the following change of potential \be
U(r)\to U(r)+\frac{{s}^2}{2Gr^2}, \label{rep(1)}
 \ee
 where $G(r)$ defines the conformal flat metric of the configuration space, $ds^2=G(r)(d{\bf r})^2$.
 Particularly, the form of classical trajectory of the system,
$\varphi =\varphi(r)$ unaffected by this kind of replacement,
whereas the orbital plane will not be orthogonal to the orbital
momentum. Instead, the angle between them will be appeared which
is defined by the expression $\cos\theta= s/J $\cite{lnp}. The
actual quantum-mechanical spectrum of the extended system
coincides with
  that of the initial one,  with minor modification of
  the possible range of orbital and magnetic quantum numbers $j,m$: $j=|s|, |s|+1,\ldots$, $m=-|s|, -|s|+1,\ldots, |s|$
  \cite{mar06}.\\

        The problem of particle moving in the field of two Coulomb
    centers (or two-center Kepler problem) was solved in  the
    middle of XIX century by Jacobi. He established the
    integrability of two-center Kepler system and of its limit
    when one of the forced centers is placed at infinity which
    yields the homogeneous potential field (we shall refer it as Kepler-Stark system) in elliptic and
    parabolic coordinates respectively. However, the
    generalization of this picture to case where Coulomb centers
    are replaced by dyons has been lacking until now.
    In our recent paper with S.Krivonos \cite{MICZ} we proposed the  generalization of the
    MICZ--Kepler replacement
    (\ref{rep(1)}) which can be used in case of $N$ Dirac monopoles,
 \be
 U\left({\bf r}\right)\rightarrow \frac{1}{2G}\left( \frac{s_1}{r_1}+...+\frac{s_n}{r_n}
 \right)^2+U\left({\bf r} \right).\label{multMICZ}
 \ee
where $s_i=eg_i$, with  $g_i$ be the magnetic charge of the $i$-th monopole located at the point  with coordinates
${\bf a}_i$, and $r_i=|{\bf r}-{\bf a}|$.

This replacement has the following  important features:
\begin{itemize}
\item The  system (without monopoles) admitting separation of
variables in elliptic/parabolic coordinates
results in the separable two-center
 MICZ-system (\ref{multMICZ}) with the Dirac monopoles placed at the foci
 of elliptic/parabolic coordinates.
\item
The system  admits the  ${\cal N}=4$ supersymmetric extension at
the following choice of  potential \be
U_{0}=\frac{\kappa}{G}\left(\sum_{I}\frac{g_I}{|{\bf r}-{\bf
a}_I|} + {\bf B}_0{\cdot\;\bf r}\right)+\frac{\kappa^2}{2G}.
\label{sup}\ee
The corresponding supersymmetric system was
constructed in earlier paper by Ivanov and Lechtenfeld \cite{IL}.
\end{itemize}
On the Euclidean case the generalized MICZ-Kepler system yields
the following generalization of the multi-center Kepler problem
describing the motion of the particle with electric charge $e$ in
the field of $N$ static Dirac dyons \be {\mathcal{H}}=\frac{1}{2}
\left(\mathbf{p}-\sum_{i=1}^N eg_i
\mathbf{A}_D\left(\mathbf{r}-\mathbf{a}_i \right)
\right)^2+\frac{1}{2}\left(\sum_{i=1}^N\frac{e
g_i}{|\mathbf{r}-\mathbf{a}_i|} \right)^2+\sum_{i=1}^N\frac{e q_i
}{|\mathbf{r}-\mathbf{a}_i|}, \label{ham1}
 \ee
 where
 \be \mathbf{A}_D\left(\mathbf{r} \right)=
 \frac{\mathbf{r}\times\mathbf{n}}{r(r-\mathbf{n}\mathbf{r})}
 \label{A}
 \ee
 is the vector-potential of the  Dirac monopole, $g_i (q_i)$  are the
 its
 magnetic(electric) charge  of  $i$-th dyon, and $\mathbf{n}_i$ is the unit vector
 pointed along the Dirac singularity line.

Since the two-center Kepler/Kepler-Stark) system admit
 separation of variables in elliptic/parabolic coordinates, the expression mentioned above provides us with integrable
  two-center  MICZ-Kepler system. Moreover, when the background dyons   have
  the same ratio of the electric and magnetic charges
  (i.e. obey trivial Dirac-Schwinger-Zwanziger charge quantization condition), the potential of the
  system (\ref{ham1}) belongs to the class (\ref{sup}), i.e. the system allows ${\cal N}=4$ supersymmetric extension.
  It is also noteworthy that the class
   of integrable models constructed in Ref.
   \cite{GW} include such  two-center MICZ-Kepler system
   in flat space  as the limiting case.
However, even  classical trajectories of the suggested
two-center MICZ-Kepler systems were not properly studied in the mentioned paper.
  Another disadvantage of the substitution (\ref{multMICZ}) is that it is ill-defined for the
   MICZ-Kepler systems on spheres and hyperboloids. Namely,
   extending in this way the (one- and two-center) Kepler system on the sphere and
   hyperboloid one will obtain the system which  \'a priori is not
   separable in  the same coordinate systems as the initial one.
   Also these systems do not belong to the class allowing the ${\cal N}=4$
   supersymmetric extensions.

In this paper we discuss these issues.

 In the {\sl Second  Section} we describe in detail the two-center MICZ-Kepler
and MICZ-Kepler-Stark systems and present their classical
solutions.

In the {\sl Third  Section} we present the alternative
model of the multi-center MICZ-Kepler system on three-dimensional
sphere and hyperboloid, as well as on any $SO(3)$ invariant
spaces.

 \section{Integrability of two-center MICZ-Kepler system}
 Now let us demonstrate the classical integrability of the two-center MICZ-Kepler system,
 and construct the solutions.
 Suppose, that background dyons  with magnetic(electric)
 charges $g_{1,2}(q_{1,2})$ are fixed on $z-$axis at points $(0,0,-a)$ and
 $(0,0,a)$ respectively. Choosing the appropriate gauge for the
 vector-potentials in spherical coordinates
 \be
 A_r^{1,2}=A_\theta^{1,2}=0, \quad A_\varphi^{1,2}=g_{1,2}\cos
 \theta_{1,2}, \label{pot}
 \ee
 we arrive at the following Hamiltonian:
 \be
 \mathcal{H}=\frac{p_r^2}{2}+\frac{p_\theta^2}{2r^2}+
 \frac{\left(p_\varphi-s_1 \cos \theta_1-s_2 \cos\theta_2 \right)^2}{2 r^2 \sin^2 \theta}
 +\frac{1}{2}\left( \frac{s_1}{r_1}+\frac{s_2}{r_2}
 \right)^2+\frac{eq_1}{r_1}+\frac{eq_2}{r_2}. \label{ham_two}
 \ee
%The
% corresponding geometry is shown in Fig. 1.
  In virtue of the
 relevant form of additional quadratic potential term the system
 allows separation of variables in elliptic coordinates given by the formulas:
 \be
 \xi=\frac{r_1+r_2}{2a}, \quad \eta=\frac{r_1-r_2}{2a}, \quad
 \varphi=\varphi. \label{elcoor}
 \ee
 Using the general kinematic relations resulting from the geometry
 of problem
 \begin{eqnarray}
  &&r_1 \sin \theta_1=r \sin \theta, \qquad
  r_2 \sin \theta_2=r \sin \theta,  \nonumber \\
  &&\cos(\theta_1-\theta_2)=\frac{4a^2-r_1^2-r_2^2}{2 r_1r_2},  \label{rel(7)}\\
  &&\cos\theta_1=\frac{r \cos\theta +a}{r_1}, \qquad\cos\theta_2=\frac{r \cos\theta-a}{r_2},
   \nonumber
  \end{eqnarray}
  one obtains the following expression for the Hamiltonian
  (\ref{ham_two}):
  \be
  \mathcal{H}=\frac{1}{2a^2\left(\xi^2-\eta^2
  \right)}\left((\xi^2-1)p_\xi^2+(1-\eta^2)p_\eta^2+\mathcal{V}(\xi)+\mathcal{W}(\eta)
  \right), \label{ham_elip}
  \ee
  where
  \begin{equation}
\mathcal{V}(\xi)=\frac{p_\varphi^2+s_+^2-2p_\varphi s_-
  \xi}{\xi^2-1}+2 a e q_+\xi , \qquad \mathcal{W}(\eta)=\frac{p_\varphi^2+s_-^2-2p_\varphi s_+
  \eta}{1-\eta^2}+2 a e q_-\eta .\label{VWe}
  \end{equation}
  Here $q_\pm=q_1 \pm q_2$, $s_\pm=s_1 \pm s_2$.  Now choosing for the
  generating function the expression $S=\varphi
  p_\varphi+S_1(\xi)+S_2(\eta)-Et$ we arrive at the following
  Hamilton-Jacobi equations:
   \begin{equation}
   \left(\xi^2-1 \right)\left(\frac{dS_1}{d\xi}
   \right)^2+\mathcal{V}(\xi)-2a^2 E \left(\xi^2-1 \right)=n, \qquad
   \left(1-\eta^2 \right)\left(\frac{dS_2}{d\eta}
   \right)^2+\mathcal{W}(\eta)-2a^2 E \left(1-\eta^2 \right)=-n, \label{Ham-Jac(20)}
   \end{equation}
   where $n$ is the separation constant which fixes the values of
   the corresponding constant of motion
   \be
   \mathcal{I}_e=-\frac{1}{\xi^2-\eta^2}\left( \eta^2\left((\xi^2-1)p_\xi^2+\mathcal{V}(\xi) \right)
    +\xi^2 \left((1-\eta^2)p_\eta^2-\mathcal{W}(\eta) \right) \right)
   \ee
   Then, integrating Eqs. (\ref{Ham-Jac(20)}) one obtains the generating function of the system
   \begin{eqnarray}
   S= p_\varphi \varphi  -E t +\int\sqrt{2a^2E+\frac{n-2 a e q_+\xi}{\xi^2-1}
   -\frac{p_\varphi^2+s_+^2-2 p_\varphi s_-\xi}{(\xi^2-1)^2}}d\xi+
  % \\ \nonumber
   \int\sqrt{2a^2E-\frac{n+2 a e q_-\eta}{1-\eta^2}
   -\frac{p_\varphi^2+s_-^2-2 p_\varphi s_+\eta}{(1-\eta^2)^2}}d\eta
   \end{eqnarray}

   Hence, we proved the exact solvability of classical two-center MICZ-Kepler system.
   Now, taking derivatives of the generating function by constant
   of motion, ${\partial S}/{\partial p_\varphi}$, ${\partial S}/{\partial n}$ and
   ${\partial S}/{\partial E}$ and putting them to
   constants (which we are claiming, without loss of generality, to be zero)
    one obtains the classical solution of the system.
Its trajectories are defined by the expressions
 \be
   \int \frac{d\xi}{(\xi^2-1) \sqrt{2a^2 E +\frac{n-\mathcal{V}(\xi)}{\xi^2-1}}}
   =\int \frac{d\eta}{(1-\eta^2) \sqrt{2a^2 E -\frac{n+\mathcal{W}(\eta)}{1-\eta^2}}}
   \label{s1}\ee
   \be
 \varphi=
   \int \frac{(p_\varphi-s_- \xi)d \xi}{(\xi^2-1)^2 \sqrt{2a^2 E +\frac{n-\mathcal{V}(\xi)}{\xi^2-1}}}
   +
   \int \frac{(p_\varphi-s_+ \eta)d\eta}{(1-\eta^2)^2 \sqrt{2a^2 E -\frac{n+\mathcal{W}(\eta)}{1-\eta^2}}},
     \label{s2}\ee
   The time evolution of the system is given by the expression
\be
 t=a^2\int \frac{d \xi}{ \sqrt{2a^2 E +\frac{n-\mathcal{V}(\xi)}{\xi^2-1}}}
  + a^2\int \frac{d\eta}{ \sqrt{2a^2 E -\frac{n+\mathcal{W}(\eta)}{1-\eta^2}}}.
\label{s3}\ee
   Here $t$ is the time and $ \mathcal{V}(\xi) $
   and $\mathcal{W}(\eta)$ are defined in Eqs. (\ref{VWe})

   \subsection{MICZ-Kepler-Stark(-Zeeman) system}
The system introduced above has an important limiting case.
   Putting one of the dyons at the infinity one obtains the
   Hamiltonian of the charged particle moving in the filed of one
   localized dyon with additional parallel homogeneous (constant uniform) electric and
   magnetic fields:
\begin{eqnarray}
{\cal H}=\frac{({\bf p}-e{\bf A}_D- e{\bf B\times
r}/2)^2}{2}+\frac{1}{2}\left(\frac{s}{r}+e{\bf B}{\bf r}
\right)^2+\frac{eq}{r} -e{\bf Er},\qquad {\bf E}\|{\bf B}
\end{eqnarray}
Due to the presence of  constant uniform magnetic and electric fields one
 could refer this system as MICZ-Kepler-Stark-Zeeman system.
As in the previous case, here one have the integrable system, i.
e. the system which allows separation of variables in parabolic
coordinates.

In order to show it we, at first, turn to
spherical coordinates in which the vector potential of the
homogeneous magnetic filed assumed to be in $z$-direction is \be
A_r=A_\theta=0, \quad A_\varphi=\frac{1}{2}B r^2 \sin^2 \theta \ee
thus, the Hamiltonian takes the form

\be {\mathcal{H}}=\frac{p^2_r}{2}+\frac{p_\theta^2}{2r^2}+
  \frac{\left(p_\varphi-s\cos\theta-\frac{1}{2}eBr^2\sin^2\theta
   \right)^2}{2r^2\sin^2\theta}+\frac{1}{2}\left(\frac{s}{r}+eBz\right)^2 +\frac{eq}{r}-eEz
\label{par_a} \ee
 Due to additional quadratic term arising from
 the relevant MICZ-Kepler replacement the corresponding
 Hamilton-Jacobi equation become separable in parabolic
 coordinates given by the following relations:
 \be
 \xi=r+z, \quad \eta=r-z
 \ee
The Hamiltonian then reads
\be {\mathcal{H}}=\frac{ 1}{2(\xi+\eta)}\left( 4\xi p_\xi^2+4\eta
p_\eta^2 +{\mathcal{V}}(\xi)+{\mathcal{W}}(\eta)\right)
-\frac{1}{2}p_\varphi e B, \label{par_ham} \ee where
\begin{eqnarray}
{\mathcal{V}}(\xi)=\frac{(p_\varphi+s)^2}{\xi}+3seB\xi-eE\xi^2+\frac{e^2
B^2 }{4}\xi^3+2+eq, \quad
{\mathcal{W}}(\eta)=\frac{(p_\varphi-s)^2}{\eta}-3seB\eta+eE\eta^2+\frac{e^2
B^2 }{4}\eta^3+2eq. \label{par_pot_(29)}
\end{eqnarray}
Then, taking for the generating function $S=\varphi
p_\varphi+S_1(\xi)+S_2(\eta)-(E-\frac{1}{2}p_\varphi e B)t$ one
arrive at the following Hamilton-Jacobi equations:
\begin{eqnarray}
&&4\xi\left(\frac{dS_1}{d\xi}
\right)^2+{\mathcal{V}}(\xi)-2E\xi=n, \qquad
4\eta\left(\frac{dS_2}{d\eta}
\right)^2+{\mathcal{W}}(\eta)-2E\eta=-n, \label{ham-jac-par(30)}
\end{eqnarray}
Here, us usual, $n$ is the separation constant which fixes the
values of specific constant of motion, responsible for the
separation of variables. For this quantity from Eqs.
(\ref{par_ham}) and (\ref{ham-jac-par(30)}) one obtains
 \be
 {\mathcal{I}}_p=\frac{1}{\xi+\eta}\left(4\xi\eta(p^2_\xi-p^2_\eta)
+\eta{\cal V}(\xi) -\xi{\cal W}(\eta)\right)
 \ee
The  generating function  obtained from the corresponding
Hamilton-Jacobi equations is
$$
S=\varphi p_\varphi -(E-\frac{1}{2}p_\varphi B)t +
\frac{1}{2}\int{\sqrt{2E-3seB-\frac{(p_\varphi+s)^2}{\xi^2}+\frac{n-2eq}{\xi}+eE\xi-
\frac{e^2B^2}{4}\xi^2}}d \xi+
$$
\be
+\frac{1}{2}\int{\sqrt{2E+3seB-\frac{(p_\varphi-s)^2}{\eta^2}-\frac{n+2eq}{\eta}-eE\eta-
\frac{e^2B^2}{4}\eta^2}}d \eta.
\ee
Hence, we prove the classical integrability of the MICZ-Kepler-Stark-Zeeman system

From the generating function  we immediately get the expressions for the trajectories of system
\be
\int \frac{d \xi}{\xi \sqrt{2 E +\frac{n-\mathcal{V}(\xi)}{\xi}}}
   =\int \frac{d\eta}{\eta \sqrt{2 E -\frac{n+\mathcal{W}(\eta)}{\eta}}},
   \label{emp1}\ee
   \be \varphi=\frac{1}{2}Bt+\frac{1}{2}\int \frac{(p_\varphi+s)d\xi}{\xi
\sqrt{2 E +\frac{n-\mathcal{V}(\xi)}{\xi}}}
   +\frac{1}{2}\int \frac{(p_\varphi-s)d\eta}{\eta \sqrt{2 E -\frac{n+\mathcal{W}(\eta)}{\eta}}},
     \label{emp2}\ee
and for the time evolution
\be
t=\int \frac{d\xi}{ \sqrt{2 E +\frac{n-\mathcal{V}(\xi)}{\xi}}}
   +\int \frac{d\eta}{ \sqrt{2 E -\frac{n+\mathcal{W}(\eta)}{\eta}}}\label{emp3}
\ee
Here $t$ is the time and $ \mathcal{V}(\xi) $
   and $\mathcal{W}(\eta)$ are defined in Eqs.(\ref{par_pot_(29)})

\section{Multi-center MICZ-Kepler system on curved space}
As was  mentioned in Introduction, any system (on conformal-flat space) without monopoles admitting the separation of variables in
elliptic/parabolic coordinates remains separable in the same coordinate systems under incorpration of monopoles supplied by the
corresponding modification of potential term given by (\ref{multMICZ}). The two-center MICZ-Kepler systems considered in previous
Section
are particular examples of these systems. However, this procedure fails
in case of the Coulomb systems on spheres and pseudospheres (two-sheet hyperboloids).
Due to the existence of hidden symmetries given by Runge-Lenz vector the Coulomb system (on  Euclidean space) admits
the separation of variables in spherical, elliptic and parabolic
coordinate.
 The Coulomb system on (pseudo)spheres also has the hidden symmetries given by the analog of Runge-Lenz vector.
Analogously  to Euclidean case, this hidden symmetry is connected
to the the separation  variables in few coordinate systems. These coordinate systems turn to the
spherical, elliptic and parabolic ones at the planar limit.
 The connection  between   spherical and Cartesian  coordinates on the sphere
  is identical  with that on the Euclidean space, whereas the discrepancy is appeared for the elliptic
  and parabolic ones (see, e.g., \cite{bogush} and refs therein).
  Therefore, the statement of the previous Section concerning
 the separability of variables for  ``MICZ-extended"  systems is no more valid for the Coulomb systems  on (pseudo)sphere.
  Moreover, it seems that  the substitution (\ref{multMICZ})
is ill-defined  on non-Euclidean spaces including (pseudo)spheres. The evidence of it stems out from the following.
When the magnetic and electric charges of the background dyons  obey trivial
 Dirac-Schwinger-Zwanziger
quantization condition  \be
g_i q_j - g_j q_i=0, %\quad n_{i j}\in\mathbb{Z}
\label{DSZ}\ee
the potential of the Euclidean multi-center MICZ-Kepler system belongs to the class (\ref{sup}) (up to unessential constant),
i.e. it admits ${\cal N}=4$ supersymmetric extension.
While the potential of the multi-center Coulomb  system on (pseudo)sphere,
\be
U({\bf r})=e\phi_{q_1,\ldots, q_N}=e\sum_{i=1}^N q_i\phi (r_i),\qquad{\rm where}\qquad \phi (r)=
\frac{1}{2r_0}\frac{1-\epsilon { r}^2}{{ r}},\quad r_i=|{\bf r}-{\bf a}_i|
\label{sC}\ee
 does not belong to the class (\ref{sup}). Here $\epsilon=1$ correspond to the sphere,
  and $\epsilon=-1$ to the pseudosphere,
and the metrics looks as follows
\be
ds^2=\frac{4r^2_0(d{\bf r})^2}{(1+\epsilon r^2)^2}. \label{met}
\ee
The (one-center) Coulomb potentials on Euclidean space, sphere and pseudosphere are nothing else, but the $so(3)$
invariant Green functions for the Laplasians defined by the corresponding metrics, $\Delta \phi_{C}=\delta({\bf r})$.
While the vector potential of the Dirac monopole is just a one-form  dual to this Green function.
Explicitly,
\be
\ast d{A}_D =-d\phi_C,\quad \Rightarrow \quad\Delta\phi_C=\delta({\bf r}),\quad \Delta\equiv\ast d\ast d+ d\ast d\ast
\label{hodge}\ee
Taking into account the duality between vectors and one- and two-forms in the three-dimensional spaces,
one can write down these expressions in the following way
 \begin{eqnarray}
-\mathbf{\nabla}\phi_C=\mbox{rot} \mathbf{A}_{D}, \quad\Rightarrow\quad \Delta\phi_C =\delta({\bf r}).
\label{vmonop}
  \end{eqnarray}
Let us write down the explicit $so(3)$ invariant solutions of this equations  on the $so(3)$ invariant space with the metrics
$d s^2=G(r)(d{\bf r})^2$. The  (zero-, one and two-)form fields are independent on the metric, hence, one
 has to choose
\be
d{\bf A}_D({\bf r})=\frac{({\bf r}\times d{\bf r})\wedge d{\bf r}}{2r^3},\qquad
d\phi(r)=\frac{d\phi(r)}{dr}\frac{{\bf r}d{\bf r}}{r}.
\ee
Taking into account the conformal flatness of the metrics, we get, from (\ref{hodge})
\be
\frac{d\phi_C}{dr}=-\frac{1}{r^2\sqrt{G(r)}},\qquad \Rightarrow\qquad \phi_C=-\int\frac{dr}{r^2\sqrt{G(r)}}.
\label{GC}\ee
Particularly, for the Euclidean space one obtains  $\phi_C=1/r$, and for the (pseudo)sphere (\ref{met}) the potential
 given in (\ref{sC}).

Now, we define the multi-center MICZ-Kepler system on the $so(3)$ invariant space by the expression
\be
  \mathcal{H}=\frac{(\mathbf{p}-e \mathbf{A}_{g_1,\ldots,g_N})^2}{2G}+\frac{e^2
  \phi^2_{g_1,\ldots, g_N}}{2}+e\phi_{q_1,\ldots,q_N}, \label{iv1}
\ee
where $\mathbf{A}_{g_1,\ldots,g_N}=\sum_i g_i\mathbf{A}_D({\bf r}-{\bf a}_i)$, $\phi_{q_1,\ldots,q_N}=\sum_i q_i\phi_C(r_i)$.

 The corresponding one-center Hamiltonian on the sphere and two-sheet hyperboloid coincides with the MICZ-Kepler
 Hamiltonian constructed within the rule (\ref{rep(1)}) up to unessential constant $\epsilon s^2/4r^2_0$.
 Hence, it is separable in the spherical coordinates, and in the modified
  elliptic and parabolic coordinates considered in Ref. \cite{bogush}.
  It seems, that two-center MICZ-Kepler systems on
spheres and pseudospheres are also separable in these coordinates.

When the electric and magnetic charges of background dyons obey the condition (\ref{DSZ}), or,
equivalently, $q_i/g_i =e\kappa $, this Hamiltonian can be represented in the form
\be
  \mathcal{H}=\frac{(\mathbf{p}-e \mathbf{A}_{g_1,\ldots,g_N})^2}{2G}+\frac{e^2(\phi_{g_1,\ldots, g_N}+\kappa/e)^2}{2}
  -\frac{\kappa^2}{2} \label{iv2}.
\ee
where ${\rm grad} (\phi_{g_1,\ldots, g_N}+\kappa/e)=-{\rm rot} \mathbf{A}_{g_1,\ldots,g_N}$.
This Hamiltonian admits the ${\cal N}=4$ supersymmetric extension on the Euclidean space (\cite{IL})
 and on the sphere (\cite{BK}) (provided  the unessential constant $\kappa^2/2$ is omitted).
In our knowledge, the  supersymmetric extensions of this systems on the generic conformal-flat space are unknown.
However, we believe, that with the explicit component expressions for the supercharges of the system on sphere \cite{BN},
we will be able to guess the  supersymmetric system on any conformal flat case.\\
%\section{Conclusion}

{\large Acknowledgements.} We are indebted  to Sergey Krivonos for collaboration at  the earlier stage of the work and drawing
our attention to his paper  on supersymmetric mechanics with monopole on sphere \cite{BK}, which allowed us to suggest the
MICZ-Kepler system un curved spaces given by Hamiltonian (\ref{iv1}). We are grateful
to George Pogosyan for pointing out us the fact that the Coulomb potential on the sphere is the
Green function of the corresponding Laplase operator,  to Tigran Hakobyan and
Vagharshak Mkhitaryan for  discussions and the
interest in work, and to the Organizers of Conference on Classical
and Quantum Integrable Systems for kind invitation and hospitality
in Dubna.
 This work was supported in part   by the grants
 NFSAT-CRDF  ARP1-3228-YE-04
 and  INTAS-05-7928.

\end{document}